\documentclass[twocolumn,a4]{aa}
\newcommand{\mr}{\mathrm}
\usepackage{graphicx,txfonts,natbib,verbatim,hyperref}
\bibpunct{(}{)}{;}{a}{}{,}

\begin{document}

\title{Effect of turbulent density-fluctuations on wave-particle
interactions and solar flare X-ray spectra}

\author{I. G. Hannah\inst{1}, E. P. Kontar\inst{1} \and H. A. S. Reid \inst{2}}

\offprints{Hannah \email{iain.hannah@glasgow.ac.uk}}

\institute{$^{1}$SUPA School of Physics \& Astronomy, University of Glasgow,
Glasgow, G12 8QQ, UK\\
$^{2}$LESIA, Observatoire de Paris, CNRS, UPMC, Universit\'{e} Paris-Diderot, 5
place Jules Janssen, 92195 Meudon Cedex, France}

\date{Received ; Accepted }

\abstract{}{The aim of this paper is to demonstrate the effect of turbulent
background density fluctuations on flare-accelerated electron transport in the
solar corona.} {Using the quasi-linear approximation, we numerically simulated
the propagation of a beam of accelerated electrons from the solar corona to the
chromosphere, including the self-consistent response of the inhomogeneous
background plasma in the form of Langmuir waves. We calculated the X-ray
spectrum from these simulations using the bremsstrahlung cross-section and
fitted the footpoint spectrum using the collisional ``thick-target'' model, a
standard approach adopted in observational studies.}{We find that the
interaction of the Langmuir waves with the background electron density gradient
shifts the waves to a higher phase velocity where they then resonate with higher
velocity electrons. The consequence is that some of the electrons are shifted to
higher energies, producing more high-energy X-rays than expected if the density
inhomogeneity is not considered. We find that the level of energy gain is
strongly dependent on the initial electron beam density at higher energy and the
magnitude of the density gradient in the background plasma. The most significant
gains are for steep (soft) spectra that initially had few electrons  at higher
energies. If the X-ray spectrum of the simulated footpoint emission are fitted
with the standard ``thick-target'' model (as is routinely done with RHESSI
observations) some simulation scenarios produce more than an order-of-magnitude
overestimate of the number of electrons $>50$keV in the source coronal distribution.}{}

\keywords{Sun:Corona -- Sun:Flares -- Sun: X-rays, gamma rays}

\titlerunning{Effect of density fluctuations on flare HXR spectra}

\authorrunning{Hannah et al.}

\maketitle

\section{Introduction}

The unprecedented RHESSI observations of solar flare hard X-rays (HXRs,
typically $> 20$keV) has forced us to consider mechanisms in addition to the
traditional collisional view of coronal electron transport. This standard
approach is of an assumed power-law of electrons above a low-energy cut-off that
propagate downwards, losing energy through Coulomb collision with the ``cold''
background plasma (whose energy is considerably lower than that of the electrons
in the beam).
The electrons will eventually stop once they have reached the higher density
chromosphere (the ``thick-target''), emitting X-rays via bremsstrahlung and
heating the plasma. The ``cold thick-target'' model CTTM
\citep{brown1971,1972SvA....16..273S} has proved popular because it provides a
straightforward relationship between the bright X-ray emission from the
chromospheric footpoints and the source coronal electron distribution. However,
many RHESSI observations are not consistent with the CTTM or produce challenging
results \citep{2011SSRv..159..107H,2011SSRv..159..301K}, which demonstrates
that essential physics is missing from the standard model.

One aspect missing from the CTTM are non-collisional processes such as
wave-particle interactions. The self-consistent generation of Langmuir waves by the electron
beam is one such process that is thought to be the related to the decimetric
radio emission in reverse-slope Type III bursts in some flares
\citep{1975Natur.258..693T,1995ApJ...455..347A,1997A&A...320..612K,
1997ApJ...480..825A}. We have previously shown that including Langmuir
waves helps in alleviating the discrepancies between the CTTM and RHESSI
observations.
For instance, the CTTM predicts a ``dip'' to appear in the flare electron
spectrum between the thermal component and just before the turnover in the electron beam
spectrum. Observationally this has not been confirmed and we showed that the
growth of Langmuir waves flattens the electron spectrum at lower
energies, maintaining a negative gradient between the thermal and non-thermal
spectral component \citep{2009ApJ...707L..45H}.  Another observational challenge
is the difference in the HXR spectral indices found between the footpoint and
coronal sources, which the CTTM predicts to be $\Delta\gamma=2$, yet RHESSI's
imaging spectroscopy of some flares has found values with $\Delta\gamma >2$
\citep{2003ApJ...595L.107E,2007A&A...466..713B,2008SoPh..250...53S,
2008A&A...487..337B,2009ApJ...705.1584S}. We found that the flattening of the
electron distribution as it propagates from the corona to the chromosphere due
to the generation of Langmuir waves can produce the observed larger differences
\citep{2011A&A...529A.109H}.

Wave-particle interactions can produce some observational features of
flares better than the CTTM but the flattening of the electron distribution to
lower energies through Langmuir wave growth produces far fainter HXR emission.
This means that a higher number of electrons need to be accelerated in the corona for
the simulations including wave-particle interactions to produce a similar
magnitude of HXRs to the standard collisional approach. Compounding this problem
further is that the CTTM itself needs a large number of electrons to be
accelerated in the corona, which conflicts with the maximum resupply rate of
electrons in the coronal acceleration region in some models.

The Langmuir waves themselves might provide a solution to this problem as they
can be scattered or refracted when interacting with an inhomogeneous background
plasma \citep{1969JETP...30..131R}, which can result in electron
acceleration \citep[e.g.][]{1989SoPh..123..343M,2001SoPh..202..131K,
2001A&A...375..629K,2010ApJ...721..864R} and increased X-ray emission
\citep{2012A&A...539A..43K}. The core idea is that the waves can be shifted to
a lower wavenumber (higher phase velocity) by interacting with the
density gradient in the background plasma, which then resonates with electrons
at higher velocity. Although this can happen in the opposite direction (with the
waves shifted to higher wavenumber), the falling power-law electron distribution
always means that this effect has the strongest consequences  for the
re-acceleration of electrons to higher energies. Recently, the role of the
non-uniform plasma has been studied for an interplanetary electron beam
\citep{2010ApJ...721..864R,2012SoPh..tmp..109R} and in the stationary (no
spatial evolution) case for solar flares in the corona
\citep{2012A&A...539A..43K}. It was found that this can lead to additional
electron acceleration. In these studies different forms of the inhomogeneity in
the background plasma were considered, including a Kolmogorov-type power-density
spectrum of fluctuations, which imitates the spectra expected from
low-frequency MHD-turbulence.

In this paper, we demonstrate the consequences of this self-consistent treatment
of electron beam-driven Langmuir waves propagating through an inhomogeneous
background plasma. This is simulated using the quasi-linear weak-turbulence
approach and is detailed in \S\ref{sec:sim}. The resulting electron and spectral
wave density distributions for a variety of forms of the input electron beam are
shown in \S\ref{sec:res}. The mean electron (deducible from observations) and
X-ray spectra are obtained for these simulations and the latter are fitted as if
they were observations, using the standard CTTM approach. This allows us to
determine the discrepancy between the CTTM-derived and true properties of the
source electron distribution in \S\ref{sec:resfit}.

\section{Electron beam simulation}\label{sec:sim}

Following the previously adopted approach
\citep{2009ApJ...707L..45H,2011A&A...529A.109H}, we simulated a 1D velocity
($v\approx v_{||} \gg v_\bot$) electron beam $f(v,x,t)$ [electrons cm$^{-4}$ s]
from the corona to the chromosphere, that self-consistently drives Langmuir
waves (of spectral energy density $W(v,x,t)$ [erg cm$^{-2}$]).
This  weakly turbulent description of quasi-linear relaxation
\citep{VedenovVelikhov1963,DrummondPines1964, 1969JETP...30..131R,hamilton1987,
2001SoPh..202..131K,2009ApJ...707L..45H} is given by

\begin{equation}\label{eq:f} \frac{\partial f}{\partial t} +v\frac{\partial
f}{\partial x}= \frac{4\pi^2 e^2}{m_\mathrm{e}^2} \frac{\partial}{\partial
v}\left( \frac{W}{v}\frac{\partial f}{\partial
v}\right)+\gamma_\mathrm{C_F}\frac{\partial}{\partial
v}\left(\frac{f}{v^2}+\frac{v_\mathrm{T}^2}{v^3}\frac{\partial f}{\partial v}
\right) \end{equation}

\begin{eqnarray}\label{eq:W} \frac{\partial W}{\partial t}+
\frac{3v_\mathrm{T}^2}{v}\frac{\partial W}{\partial x}
+\frac{v^2}{L}\frac{\partial W}{\partial v}=
 \left( \frac{\pi
\omega_\mathrm{p}}{n}v^2 \frac{\partial f}{\partial
v}-\gamma_\mathrm{C_W}-2\gamma_\mathrm{L} \right)W+Sf, \nonumber \\
\end{eqnarray}

\noindent where $n$ the background plasma density, $m_\mathrm{e}$ the electron
mass and $\omega_\mathrm{p}^2=4\pi n e^2/m_\mathrm{e}$ is the local plasma
frequency.
The first terms on the right-hand side of Eqs. (\ref{eq:f}) and (\ref{eq:W})
describe the quasi-linear interaction, the other terms the Coulomb collisions
$\gamma_\mathrm{C_F}=4\pi e^4 n\ln{\Lambda}/m_\mathrm{e}^2$ and
$\gamma_\mathrm{C_W}= \pi e^4n \ln{\Lambda}/(m_\mathrm{e}^2 v_\mathrm{T}^3)$
with $\ln{\Lambda}$ the Coulomb logarithm, Landau damping
$\gamma_\mathrm{L}=\sqrt{\pi/8}\omega_\mathrm{p}\left(v/v_\mathrm{T}
\right)^3\exp{ (-v^2/2v_\mathrm{T}^2 ) }$, and spontaneous wave emission
$S=\omega_\mathrm{p}^3m v\ln{(v/v_\mathrm{T})}/(4\pi n)$. The simulations here
feature two changes over the previous work
\citep{2009ApJ...707L..45H,2011A&A...529A.109H}. The first minor change  is the
inclusion of the diffusion in velocity space due to collisions (final term in
the brackets at the end of Eq. (\ref{eq:f})). Previously only the drag term was
used, which described a ``cold'' target situation in which the energy in the
beam electrons is considerably higher than that of the background thermal
distribution. Including the diffusion term allows a more realistic
treatment of electrons at energies closer to the thermal background, i.e.
a ``warm'' target.  The second, and more substantial, change is the inclusion
of a turbulent background plasma and the effect of this density gradient on the
plasma waves (third term on the left-hand side of Eq.
(\ref{eq:W}). This is done using the characteristic scale of the plasma
inhomogeneity $L^{-1}=(\partial \omega_\mathrm{p}/\partial
x)\omega_\mathrm{p}^{-1}=(\partial n/\partial x)(2n)^{-1}$ as used previously by
\citet{2001SoPh..202..131K} and \citet{2010ApJ...721..864R}.

The electron distribution is simulated in the velocity domain from the 1MK
background plasma thermal velocity $v_\mathrm{T}$ up to $115v_\mathrm{T}$ This
range extends below the observational capacity of RHESSI (down to about 86eV
instead of RHESSI's 3keV limit), well into the range where the thermal emission
will dominate. Although the treatment of the thermal distribution is beyond the
scope of this work, the energy range is included to demonstrate the possible
effect these lower energy electrons have on the higher energy population.

The initial electron distribution is Gaussian of width $d=2\times10^{8}$cm and a
broken power-law in velocity, which is flat below the break. This break is
effectively the low-energy cut-off. We used $v_\mathrm{C}\approx
9v_\mathrm{T}$ ($E_\mathrm{C}=7$keV), and the power-law above it has an index of
$2\delta_\mathrm{b}$ (hence a spectral index of $\delta_\mathrm{b}$ in energy
space), i.e.

\begin{equation}\label{eq:init} f(v,x,t=0)\propto  n_\mathrm{b}
\exp{\left(-\frac{x^2}{d^2}\right)} \left\{ \begin{array}{l l} 1 & \quad
\mbox{
$v_\mathrm{T}<v<v_\mr{C}$}\\ \left( v/v_\mr{C}\right)^{-2\delta_\mathrm{b}} &
\quad \mbox{ $v_\mathrm{C}\le v < v_\mathrm{0} $},\\
\end{array}\right.\end{equation}

\noindent where $v_\mathrm{0}=90v_\mathrm{T}$ is the maximum initial beam
velocity, $n_\mathrm{b}=\int_{v_\mathrm{C}}f\mathrm{d}v$ is the electron beam
density above the break/low-energy cut-off. For the simulations presented in
this paper we used a beam density above the break of
$n_\mathrm{b}=10^8$cm$^{-3}$.
At the start of the simulations this beam was instantaneously injected at a
height of 40Mm above the photosphere and was not replenished, with the spatial
grid extending from 52Mm down to 0.3Mm.

A finite difference method \citep{2001CoPhC.138..222K} is used to solve Eqs.
(\ref{eq:f}) and (\ref{eq:W}), and the code is modular which makes it
easy to consider the effects of the different processes. We consider three distinct simulation
setups within this paper, namely:
\begin{itemize}
  \item \textbf {B: beam-only:} We only consider the propagation of the electron
  distribution subject to Coulomb collisions with the background
  plasma, similar to the CTTM. Specifically, we only solve Eq. (\ref{eq:f}),
  ignoring the quasi-linear term (first term on the right-hand side).
 \item \textbf {BW: Beam and waves:} We consider the propagation of
    the electron distribution and the self-consistent driving of Langmuir waves
    but without the plasma inhomogeneity term. This is solving both Eqs.
    (\ref{eq:f}) and (\ref{eq:W}) but without the last term on the left-hand side of Eq.
    (\ref{eq:W}), i.e. ignoring $L^{-1}$.
   \item \textbf{BWI: Beam, waves and inhomogeneity:} We consider the
   propagation of the electron distribution, the self-consistent driving of
   Langmuir waves, and the wave interaction with the inhomogeneous background plasma.
    This is the full solution to both Eqs. (\ref{eq:f}) and
    (\ref{eq:W}).
\end{itemize}
\noindent These three different simulation setups allow us to investigate
electron transport due to collisions (B, akin to the CTTM), collisions and
wave-particle interactions (BW), and collisions, wave-particle interactions
and the plasma inhomogeneities (BWI).

\begin{figure}\centering \includegraphics[width=65mm]{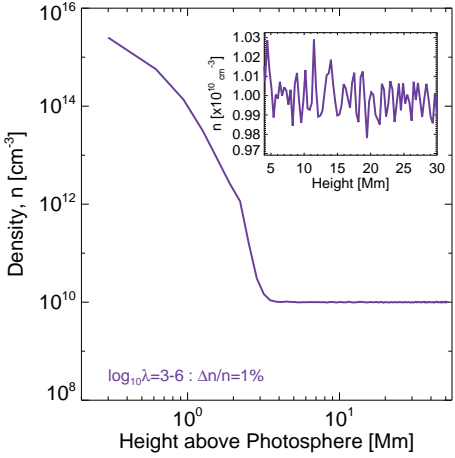}
\caption{\label{fig:none}Background plasma density profile $n$. The inset
view is of a zoomed portion of the density profile, showing that the density fluctuations (1000 of them with wavelengths
between $10^3\le\lambda_i\le 10^6$ cm and amplitude of 1\%) have been added to
the background profile.}
\end{figure}

\begin{figure}\centering \includegraphics[width=65mm]{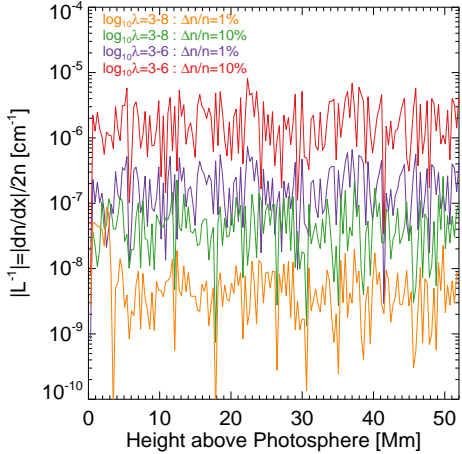}
\caption{\label{fig:lall}Magnitude of the characteristic scale of the plasma
inhomogeneity $L^{-1}$ as a function of height above the photosphere. It is
shown for two different wavelength ranges, $10^3\le\lambda_i\le 10^6$ cm and
$10^3\le\lambda_i\le 10^8$ cm, using amplitudes of 1\% and 10\%.}
\end{figure}

\subsection{Background density fluctuations}

The main component of the background density $n_0$ is constant in the corona
($10^{10}$cm$^{-3}$) and sharply rises through the transition region and
chromosphere (below 3Mm). It is shown in Figure \ref{fig:none} and was used
previously in \citet{2011A&A...529A.109H}. The additional components presented
in this paper are the density fluctuations, which are drawn from a turbulent
Kolmogorov-type $\beta=5/3$ power density spectrum, i.e.

\begin{equation}\label{eq:ndn}
n=n_0\left[ 1+C\sum_{i=1}^N \lambda_i^{\beta/2} \sin \left(\frac{2\pi
x}{\lambda_i}+\phi_i\right) \right],
\end{equation}

\noindent where $\lambda_i$ are the wavelengths of the density fluctuations, and
$C$ is a normalisation constant used to control their amplitude via
$\langle\Delta n\rangle /\langle n\rangle=(C^2\sum_i^N \lambda_i^\beta/2)^{1/2}$, as
implemented in \citet{2010ApJ...721..864R}. Here $N=1000$ fluctuations are added
to the background density profile with random phases $\phi_i$ between $0$
 and $2\pi$ and wavelengths of either $10^3\le \lambda_i \le 10^6$ cm and $10^3
 \le \lambda_i \le 10^8$cm chosen randomly in logarithmic space. We investigated
 two wavelength ranges. One extends to the Mm range, though in both cases
 it is smaller than the whole simulation region, choosing $C$ to achieve
 amplitudes of either 1\% or 10\%. The inset plot in Figure \ref{fig:none} shows the fluctuations for
 the $10^3\le \lambda_i \le 10^6$ cm and 1\% case, which are also present in the
 main plot. The most important aspect of the fluctuations is not the wavelength range
 or amplitude used, but the resulting magnitude of the density gradient, which
 influences the waves via the characteristic scale of the plasma inhomogeneity
 $L=2n(\partial n/\partial x)^{-1}$ in Eq. (\ref{eq:W}). To calculate this the
 density gradient of the fluctuations has to be analytically found so that they
 are accurately included, i.e.

\begin{eqnarray}\label{eq:ndnx}
\frac{\partial n}{\partial x}=\frac{\partial n_0}{\partial
x}+C\sum_{i=1}^N&\frac{\partial n_0}{\partial
x} \lambda_i^{\beta/2}& \sin \left(\frac{2\pi
x}{\lambda_i}+\phi_i\right) \nonumber \\
& & + 2n_0\pi\lambda_i^{\beta/2-1}\cos
\left(\frac{2\pi x}{\lambda_i}+\phi_i\right).
\end{eqnarray}

\noindent The resulting $L$ for the four different configurations of the
fluctuations is shown in Figure \ref{fig:lall}.

\subsection{Simulated X-ray and mean electron spectrum}
From these simulations we can compute the X-ray spectrum $I(\epsilon)$ and the
mean electron flux spectrum $\langle nVF(E) \rangle$, deducible from the
observed X-ray spectrum. For the X-ray spectrum $I(\epsilon)$ we used

\begin{equation}\label{eq:ph_sum}
I(\epsilon)=\frac{A}{4\pi R^2} \sum_{E}
\sum_{x}
\sum_{t}\left[n(x)\frac{f(v,x,t)}{m_\mr{e}} Q(\epsilon,E) \right]dE dx,
dt \end{equation}

\noindent where $A$ is the area of the emitting plasma (which we took to be the
square of the full width at half-maximum of the Gaussian spatial distribution,
i.e.
$8(\ln{2}d^2)$), and $Q(\epsilon,E)$ is the bremsstrahlung cross-section
\citep{1959RvMP...31..920K,1997A&A...326..417H}. We calculated the mean electron
flux spectrum $\langle nVF(E) \rangle$, which is deducible from the X-ray
spectrum \citep[e.g.][]{Brown_etal2006} by

\begin{equation}\label{eq:nvf}
\langle nVF(E) \rangle=\frac{A}{4\pi R^2} \sum_{x}
\sum_{t}\left[n(x)\frac{f(v,x,t)}{m_\mr{e}} \right]dx dt,
\end{equation}

\noindent where $F(E)\mathrm{d}E=vf(v)\mathrm{d}v$ is the electron flux spectrum
as a function of energy, not velocity. Most of the spectra shown in
\S\ref{sec:res} are summed over the whole simulation, but for the X-ray
footpoint spectrum (\S\ref{sec:resfit}) the summation is over $0.3\le x\le 3$Mm.

\begin{figure*}\centering
\includegraphics[width=160mm]{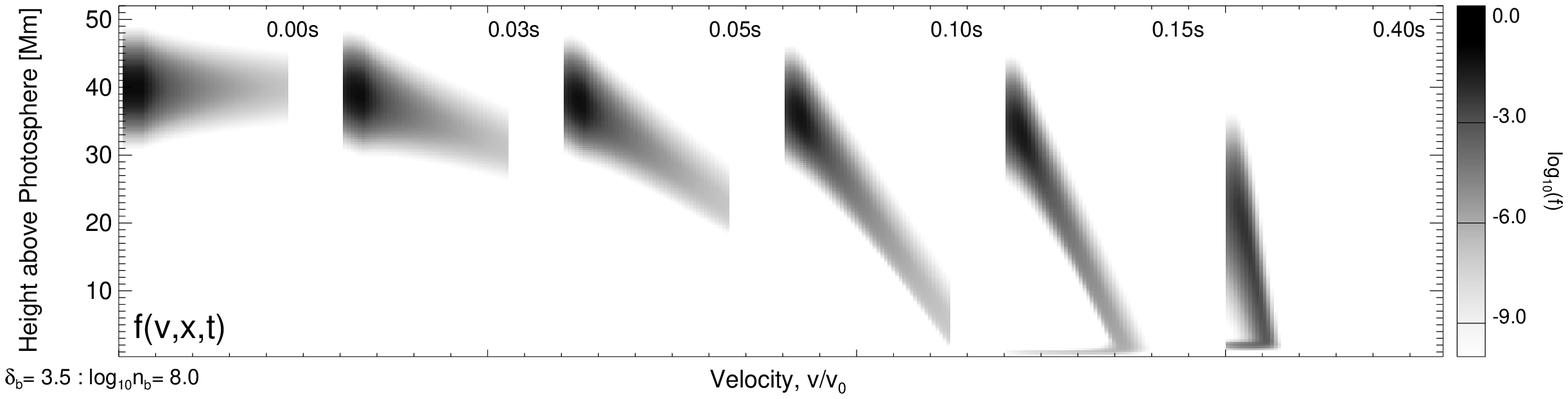}\\
\includegraphics[width=160mm]{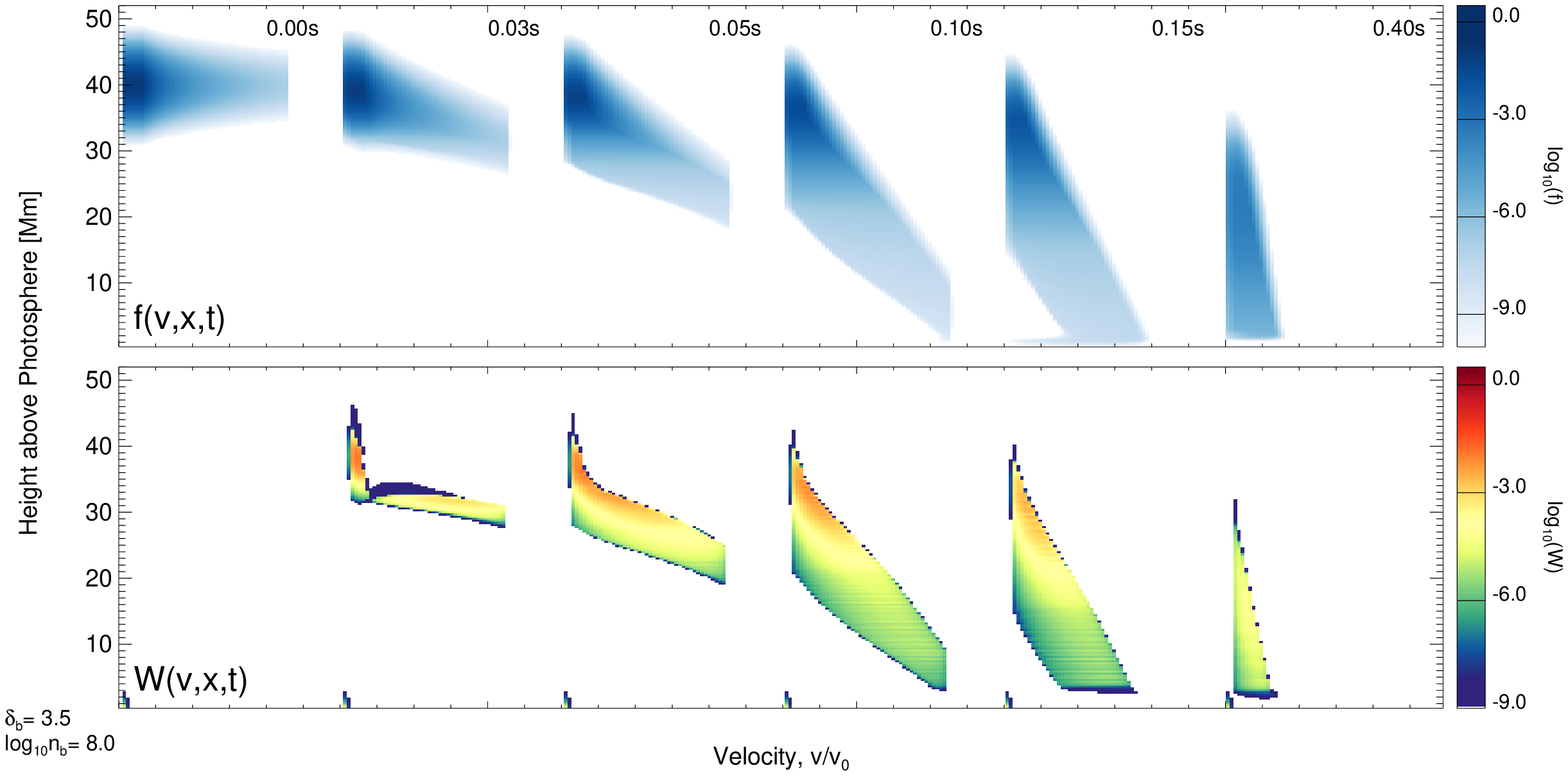}\\
\includegraphics[width=160mm]{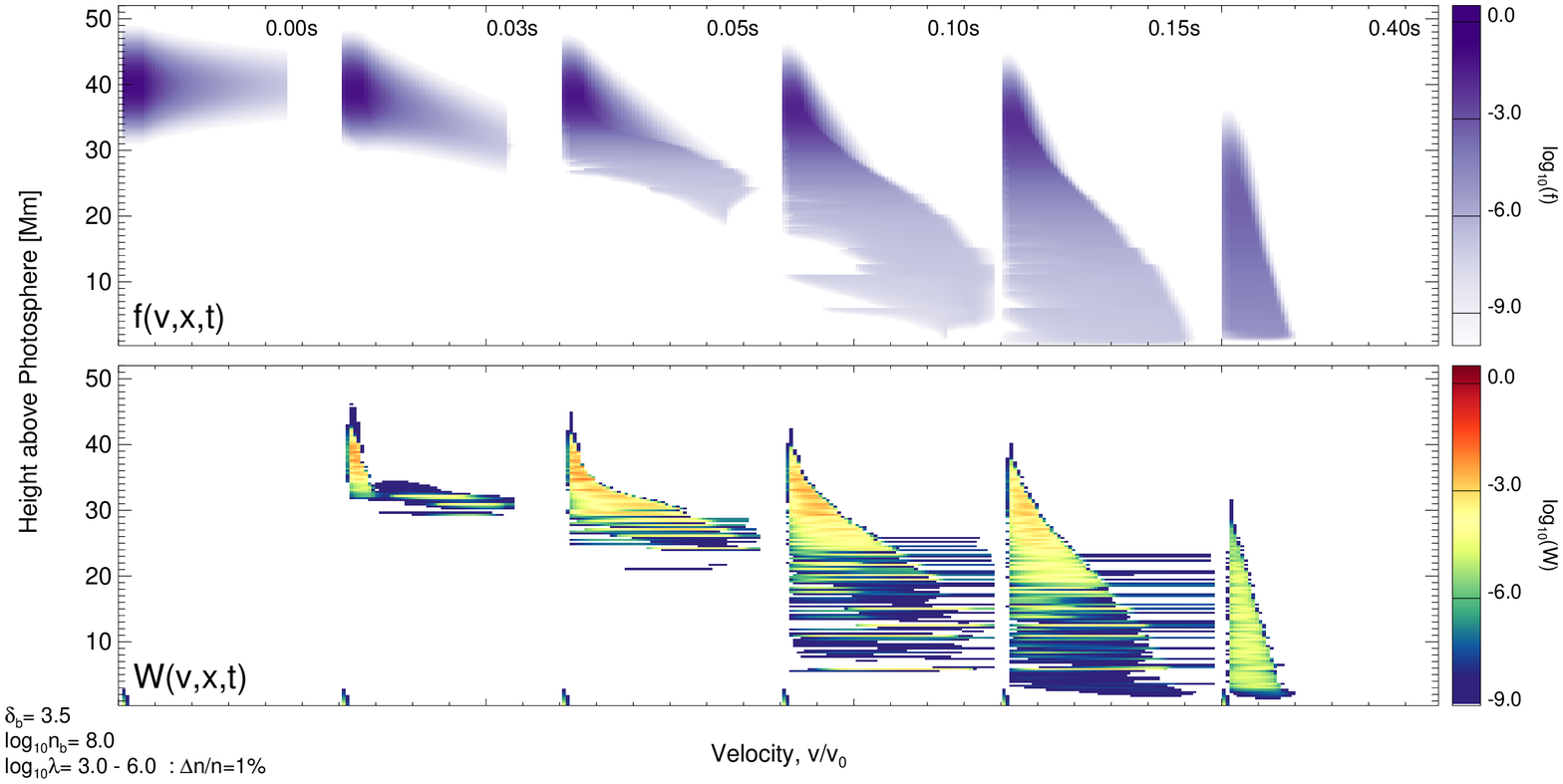}
\caption{\label{fig:fxv}Evolution of the electron $f(v,x,t)$ and wave spectral
energy distributions $W(v,x,t)$, with time increasing left to right, for
different simulation setups (top to bottom) but all with $\delta_\mathrm{b}=3.5$
and $n_b=10^8$cm$^{-3}$.
(Top) Electron distribution for the beam-only B simulation. (Middle) Electron
distribution and wave spectral energy distribution for the beam and waves BW
simulation. (Bottom) Electron distribution and wave spectral energy distribution
for the beam, waves and inhomogeneity BWI simulation.}
\end{figure*}

\section{Simulation results}\label{sec:res}

Results from one configuration of the three simulation setups is shown in Figure
\ref{fig:fxv}, all using an initial beam of $\delta_\mathrm{b}=3.5$ and
background density fluctuations of $10^3\le \lambda_i \le 10^6$ and $\Delta
n/n=1\%$. Shown here are the
electron $f(v,x,t)$ and wave spectral energy distributions $W(v,x,t)$, in terms
of velocity-vs-distance travelled in each frame with time increasing from left
to right. The first two configurations, beam-only (B, top panel) and beam and
waves (BW, middle panels), show similar results to those that we have previously
published \citep{2009ApJ...707L..45H}. Here we are using a higher beam
density, however  we have a flat (instead of no) electron distribution below
$E_\mathrm{C}$, and there are density fluctuations in the background plasma. In
the beam-only case we see that the fastest electrons reach the lower atmosphere
first, quickly lose energy to the high-density background plasma and leave the
simulation grid. The bulk of the electron distribution takes longer to lose energy through
Coulomb collisions with the background plasma. After 1~second in simulation time
the electron distribution is no longer present. When the wave-particle
interactions are included (BW, middle panel of Figure \ref{fig:fxv}), the
electron distribution immediately flattens/widens in velocity space, with
electrons shifted to lower energies. After $t=0.03$s two components of Langmuir waves have
developed: one at lower energies through the spontaneous emission, the $Sf$ term
in Eq. (\ref{eq:W}) and another across a wide range of velocities through the
propagation of the fastest electrons away from the bulk of the distribution, the
$\partial f/\partial x$ term in Eq. (\ref{eq:W}).

\begin{figure*}\centering
\includegraphics[width=42mm]{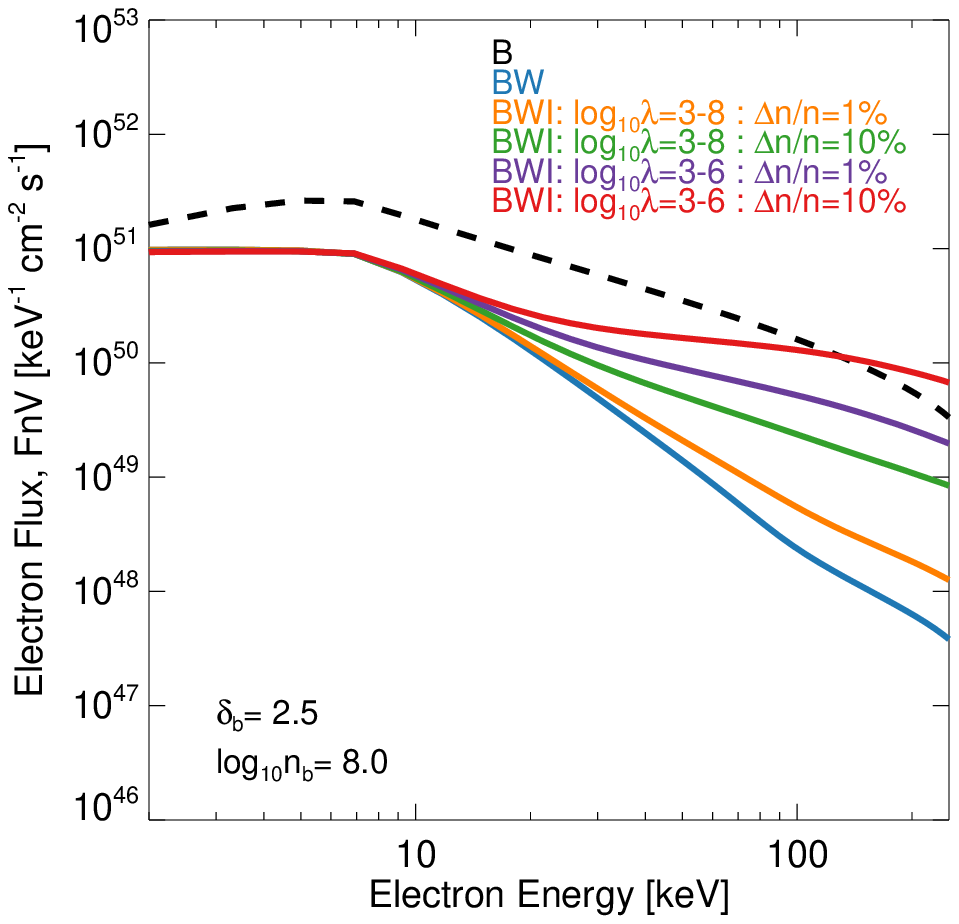}
\includegraphics[width=42mm]{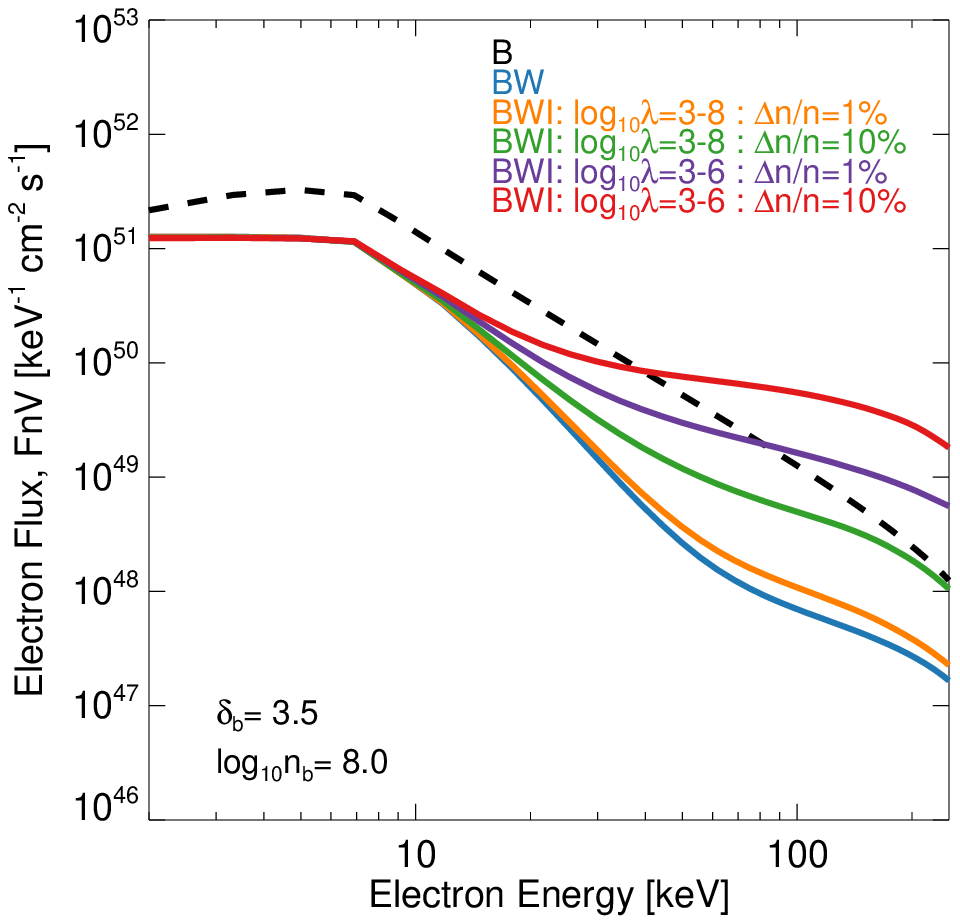}
\includegraphics[width=42mm]{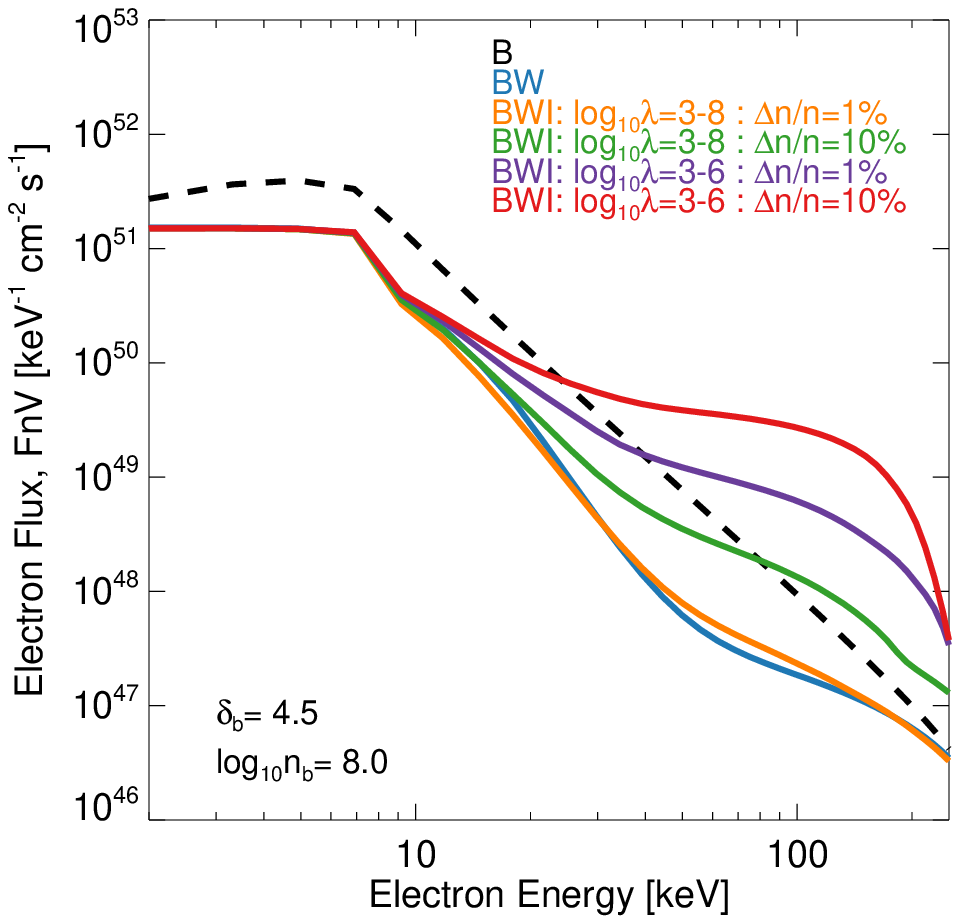}
\includegraphics[width=42mm]{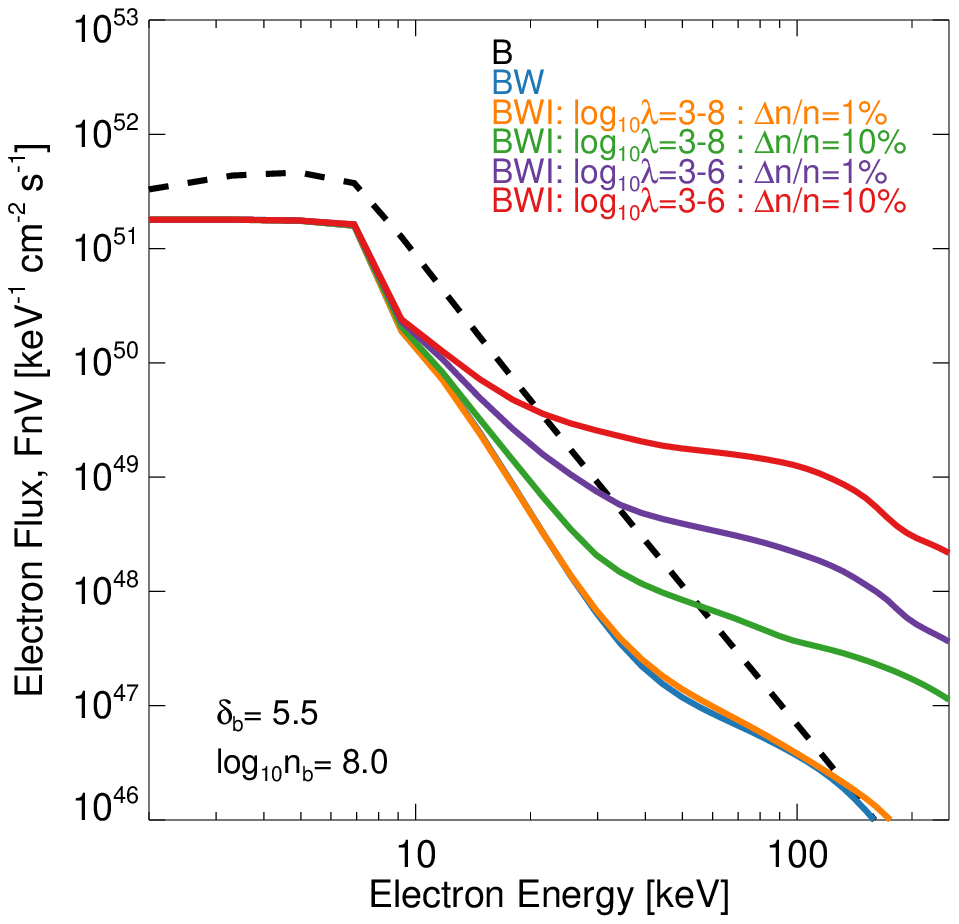}
\caption{\label{fig:fspec}Mean electron flux spectrum (spatially integrated and
time averaged) for initial electron beams of power-law index
$\delta_\mathrm{b}=2.5,3.5,4.5,$ and $5.5$ (increasing left to right). The
different colour lines indicate the different simulation setups : B (black
dashed), BW (blue), and BWI (orange, green,purple, and
red). In the last case two different wavelength ranges are used ($10^3\le
\lambda_i \le 10^6$ cm and $10^3\le \lambda_i \le 10^8$ cm) with fluctuations of
the amplitudes of 1\% and 10\%.}
\end{figure*}

\begin{figure*}\centering
\includegraphics[width=42mm]{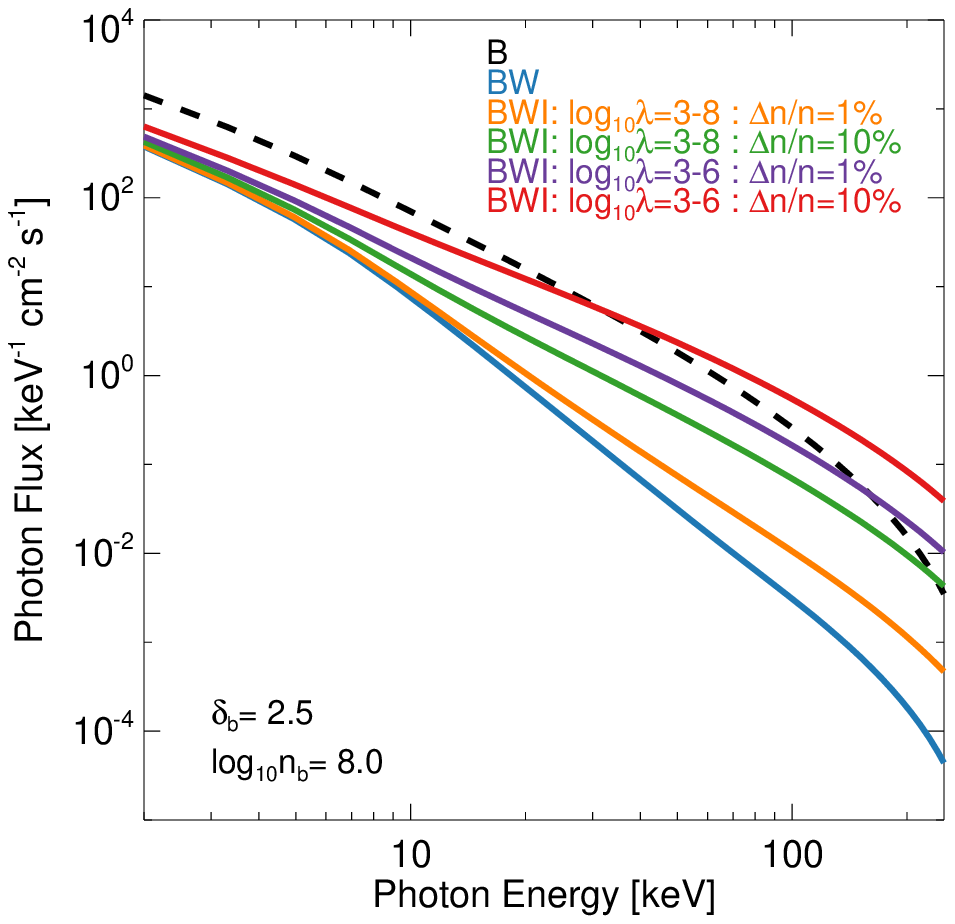}
\includegraphics[width=42mm]{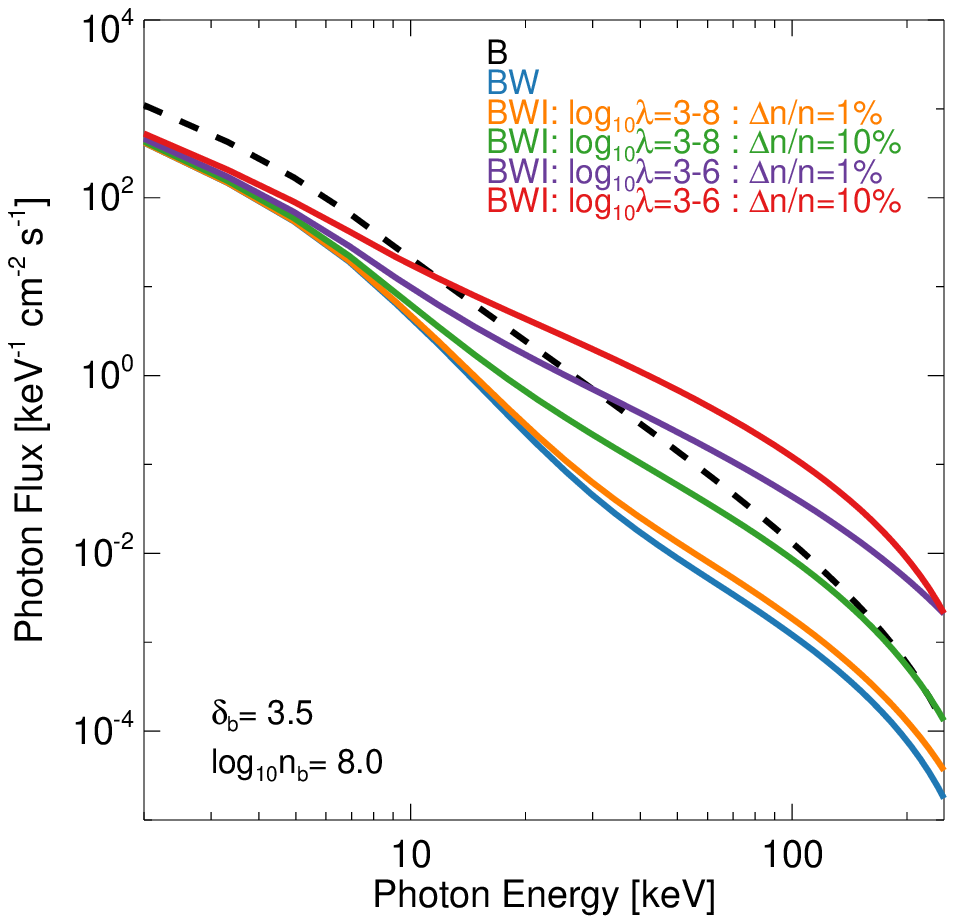}
\includegraphics[width=42mm]{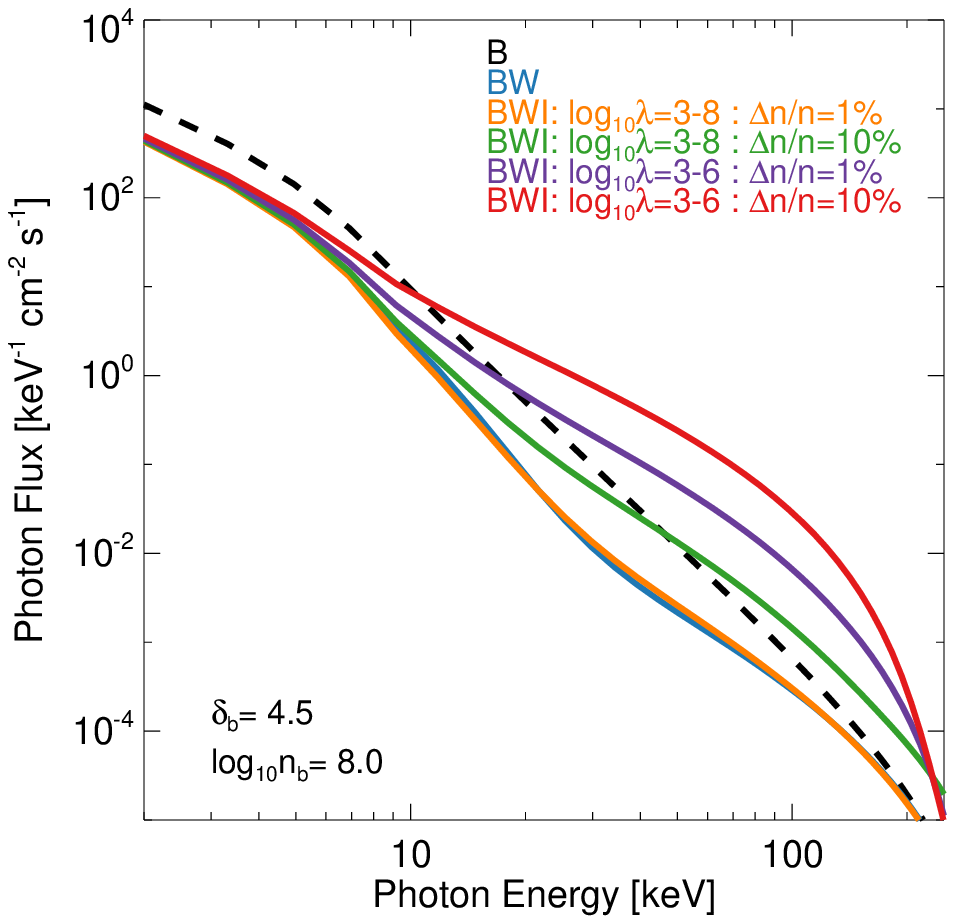}
\includegraphics[width=42mm]{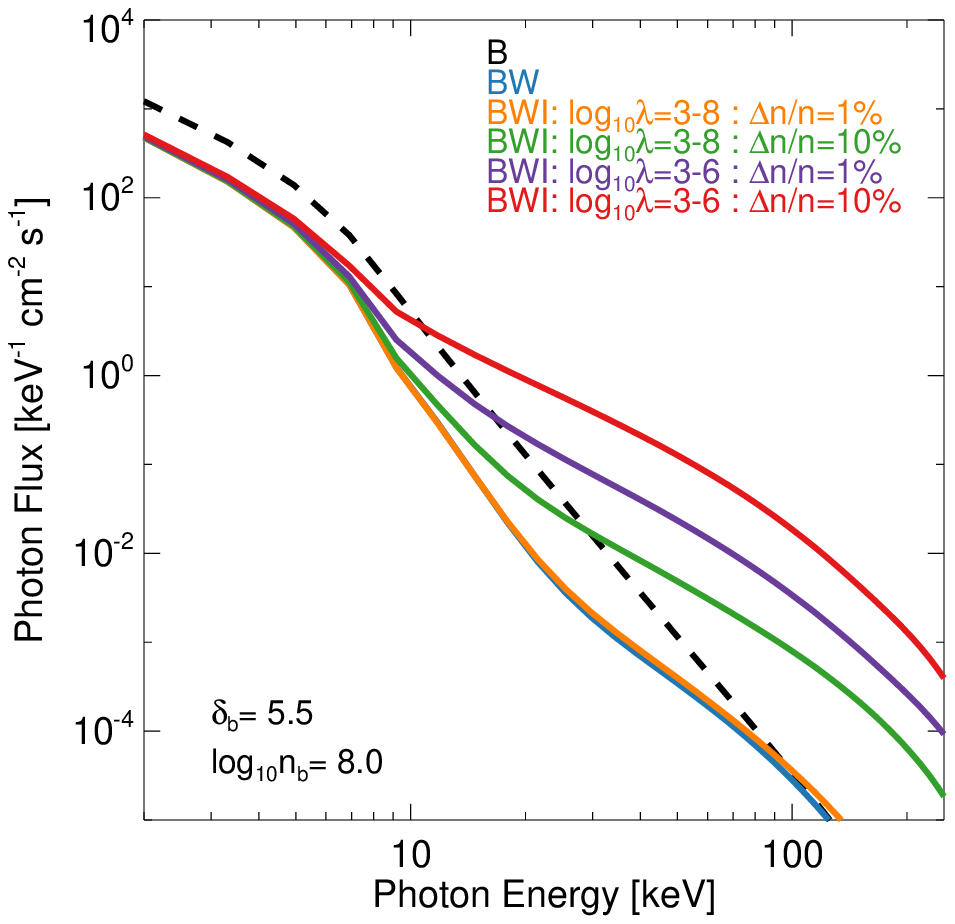}
\caption{\label{fig:xspec}X-ray spectrum (spatially integrated and
time averaged) for initial electron beams of power-law index
$\delta_\mathrm{b}=2.5,3.5,4.5,$ and $5.5$ (increasing left to right). The
different colour lines indicate the different simulation setups : B (black
dashed), BW (blue), and BWI (orange, green,purple, and
red). In the last case two different wavelength ranges are used ($10^3\le
\lambda_i \le 10^6$ cm and $10^3\le \lambda_i \le 10^8$ cm) with fluctuations of
the amplitudes of 1\% and 10\%.}
\end{figure*}

The result that these simulations produce similar results to our previous work
confirms that the density fluctuations only play a role once the inhomogeneity
term $L$ term is included in Eq.
(\ref{eq:W}), which is shown in the bottom panels of Figure (\ref{fig:fxv}). The
BWI shows a dramatic change over the other
simulations, with streaks appearing in the wave spectral density plots because
the waves shift to lower and higher phase velocity (or higher and
lower wavenumber). The effect on the electron distribution is to pull it out in
clumps across the velocity-space, which is most evident in the $t=0.10$s frame.
The leading edge of the electron distribution is clearly pushed out to higher velocities
compared to the other setups, i.e. at times $t=0.10, 0.15$s the frames of the
electron distributions of the B (black) and BW (blue) setups do not extend to
energies as high as the BWI (purple) setup in Figure \ref{fig:fxv}.

The change in energy is most evident when the spatially-integrated mean electron
spectra $\langle nVF(E) \rangle$ are calculated for the simulations, as shown in
Figure \ref{fig:fspec}. Here all configurations of the simulations are shown,
indicated by different coloured lines, and the panels show the spectral indices
of the initial distribution $\delta_\mathrm{b}=2.5,3.5,4.5,$ and $5.5$
increasing from left to right. The mean electron spectrum of  the simulations
shown in Figure \ref{fig:fxv} are shown in the second plot in Figure
\ref{fig:fspec}, the same colours are used for the electron distributions in
each different setup. For the hardest (i.e. flattest, $\delta_\mathrm{b}=2.5$,
left panel Figure \ref{fig:fspec}) initial spectrum almost all different setups
are substantially lower than for the beam-only case. Only the fluctuations with
the steepest density gradient (10\% and $10^3\le \lambda_i \le 10^6$ cm) produce
electrons at higher energies than the purely collisional setup, but this only
occurs at the highest energies ($>100$keV). With steeper initial spectral
indices (larger $\delta_\mathrm{b}$) we find that more electrons have been
accelerated with even lower levels of density fluctuations, though the
inhomogeneity needs to have $L^{-1}\geq 10^{-8}$cm$^{-1}$. This may depend on
the way the initial distributions were normalised. The same beam density above
the low-energy cut-off was used throughout
$n_\mathrm{b}=\int_{v_\mathrm{C}}f\mathrm{d}v$, but for steeper spectra this
results in more electrons at energies just above $E_\mathrm{C}$. Therefore there
is a higher number (about a factor of two from Figure \ref{fig:fspec}) of
electrons with about 10 keV with the steeper spectra available to be
re-accelerated by the shifted waves. Even for the strongest electron
acceleration caused by the density fluctuations, it is only above about 20 keV
that there are more electrons than in the beam-only setup. In all simulations
where the wave-particle interactions are present, the Langmuir wave generation
flattens the spectrum, which reduces the number of low-energy electrons compared
to the purely collisional case.

The result of the wave scattering for the HXR spectrum is more complicated since
an electron can produce an X-ray below its energy. This is because higher energy
electrons can travel farther into the dense regions of the lower solar
atmosphere,  producing substantially stronger HXR emission. The spatially
integrated X-ray spectra for the different simulation setups and configurations
are shown in Figure \ref{fig:xspec}. As with the mean electron spectrum, the
most significant changes are observed in the simulations with the softest
(steepest $\delta_\mathrm{b}=5.5$, last panel of Figure \ref{fig:xspec}) initial
electron distributions. Here the X-ray emission is up to several orders of
magnitudes higher than the beam-only case, whose spectrum is flatter down to 10
keV. This trend continues with the harder initial electron distributions
producing flatter X-ray spectrum compared to the purely collisional case.
Again with the hardest initial spectrum ($\delta_\mathrm{b}=2.5$, first panel
Figure \ref{fig:xspec}) only the strongest density fluctuations produce X-ray
emission higher than the beam-only case, but this extends to about 30 keV,
whereas in the electron spectrum it is only higher $>100$ keV.

\begin{figure}\centering \includegraphics[width=65mm]{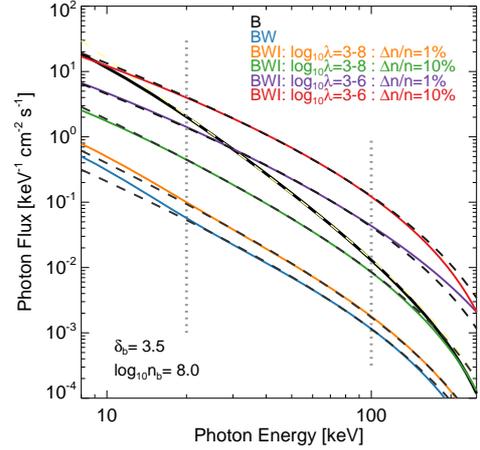}
\caption{\label{fig:specfits} Footpoint X-ray spectrum for an initial electron
power-law index of $\delta_\mathrm{b}=3.5$ and the different simulation setups
(coloured lines). Each is fitted over $20-100$ keV using the
\texttt{f\_thick2.pro} model, which is indicated by the dashed lines and is
similar to the beam-only simulation (B, black line).}
\end{figure}

\subsection{Fitting the footpoint X-ray spectrum}\label{sec:resfit}

To quantify the effect of the wave-particle interactions and density
fluctuations on the X-ray spectrum, we fitted them as if they were actual
observations. We specifically fitted the footpoint X-ray spectrum, where in Eq.
(\ref{eq:ph_sum}) we summed over the region $0.03\le x\le 3$Mm instead of the
whole simulation, because in RHESSI observations the flare spectrum is mostly
dominated by the footpoint emission from the chromosphere. These spectra were
fitted using the implementation of the CTTM in the OSPEX software
\texttt{f\_thick2.pro}, an optimised version of the routine by G. Holman (usage
examples are given in \citet{Holman2003}), available in the SolarSoft X-ray
package\footnote{\href{http://hesperia.gsfc.nasa.gov/ssw/packages/xray/}
{http://hesperia.gsfc.nasa.gov/ssw/packages/xray/}}. We fitted a single
power-law as the source spectrum, setting the low-energy cut-off
($E_\mathrm{C}=7$keV) and maximum energy to be the same as the initial electron
distribution in our simulations. The fitting can be highly sensitive to the
low-energy cut-off, so by fixing it to the true simulation value, we avoided
this problem, as well as the issue of the missing thermal component at low energies
that would be present in real spectral observations. We therefore have two free
parameters in our model fit:
the total number of electrons $N(>E_\mathrm{C})$ and the spectral index
$\delta_\mathrm{TT}$ of the source distribution. We fitted, by minimizing
$\chi^2$, the simulated footpoint spectra over 20 to 100 keV, the typical energy range
used in RHESSI observations, which also avoids complications of the thermal
component at low energies and simulation edge effects at higher energies.

The simulated footpoint spectra (colour lines) and their \texttt{f\_thick2.pro}
fits (dashed lines) for the initial spectral index $\delta_\mathrm{b}=3.5$ are
shown in Figure \ref{fig:specfits} for all simulation setups. In all cases the
model fits the simulated spectra very well over the chosen energy range
(indicated by the dotted vertical lines). The fitted \texttt{f\_thick2.pro}
model should be a reasonable match to our simulation B and we obtain
$\delta_\mathrm{TT}=3.7$ and $N(>E_\mathrm{C})=4\times10^{35}$ electrons. The
slight discrepancy between the fitted and true spectral index (3.7 vs 3.5) for
the source distribution is because the \texttt{f\_thick2.pro} model is
steady-state and stationary where as our simulation includes the time and
1D-spatial evolution of an injected (not continuous) electron beam. For the
simulations with different initial spectral indices we again find only a small
discrepancy to the fitted values, obtaining 2.8, 4.7, and 5.6. In all these
B simulations we obtained the total number of electrons in the range of
$N(>E_\mathrm{C})=3.5 - 4.7\times 10^{35}$ electrons.

\begin{figure}\centering \includegraphics[width=65mm]{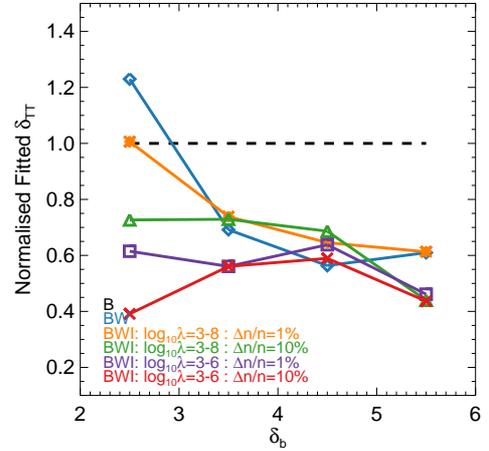}
\caption{\label{fig:delfits}Power-law index of the electron distribution
$\delta_\mathrm{TT}$ found from fitting \texttt{f\_thick2.pro} to the simulated
footpoint X-ray spectra as a function of the actual initial power-law index
$\delta_\mathrm{b}$. The index $\delta_\mathrm{TT}$ is normalised by those
found from the fit to the beam-only (B, black line) X-ray spectrum.}
\end{figure}

\begin{figure*}\centering
\includegraphics[width=50mm]{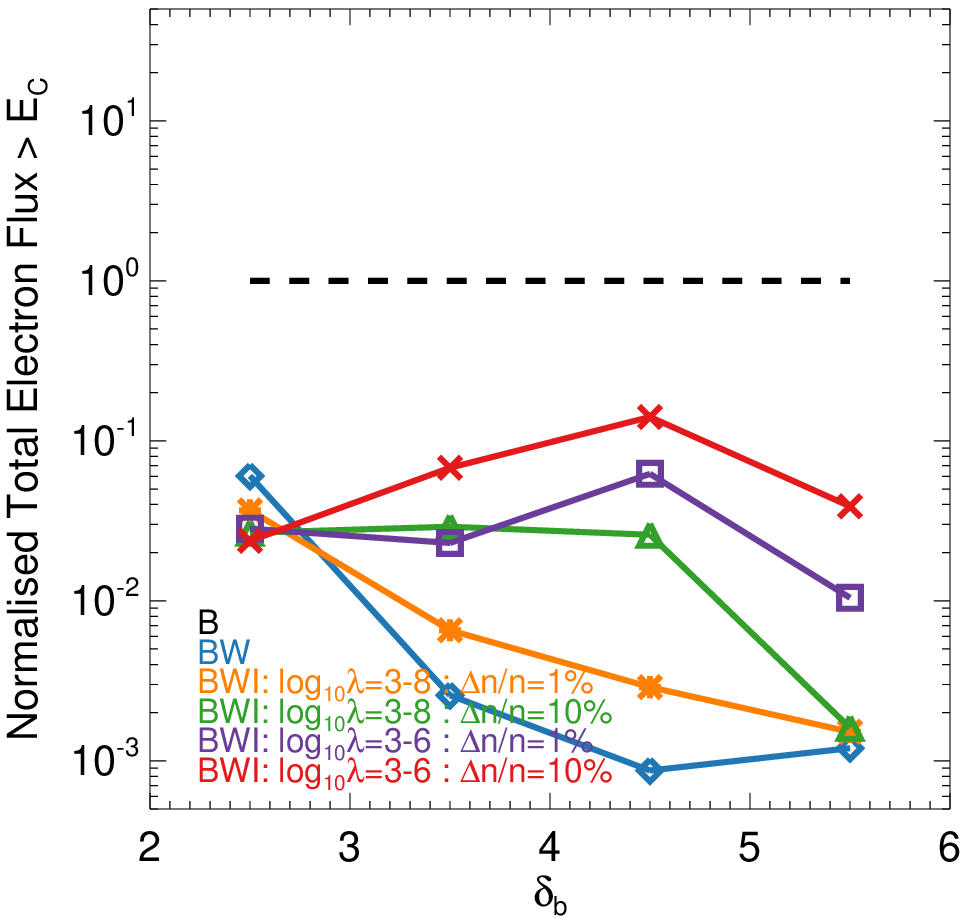}
\includegraphics[width=50mm]{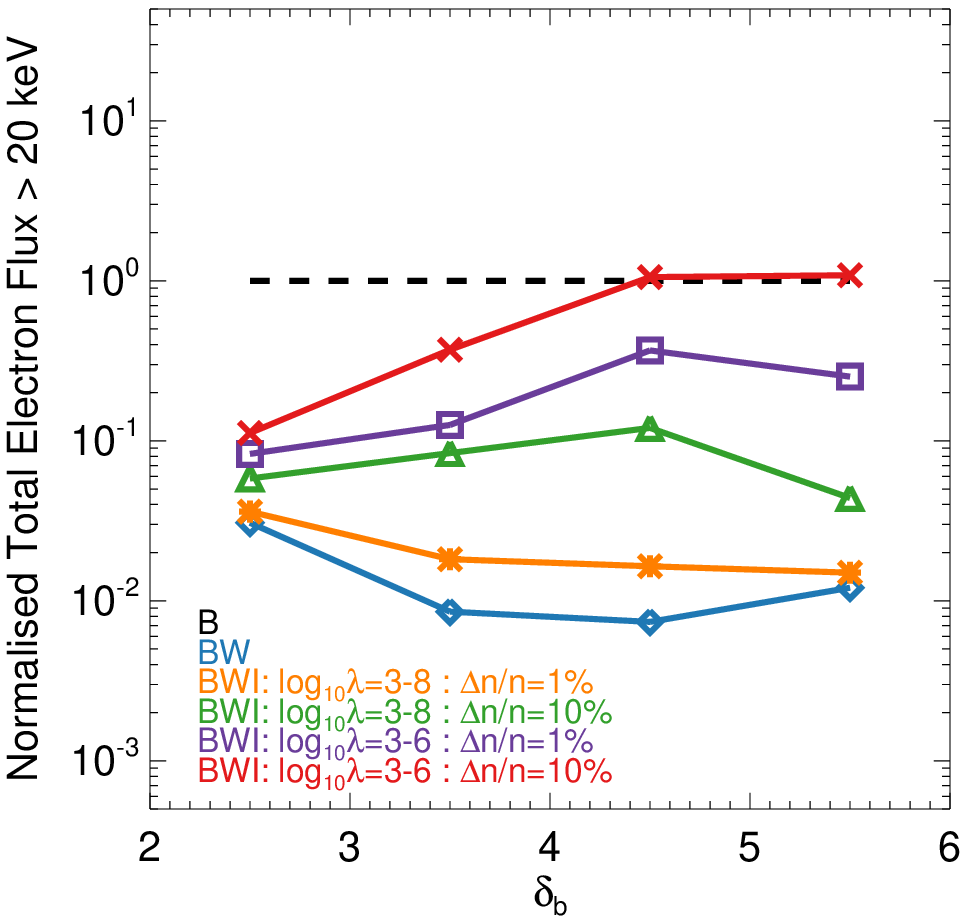}
\includegraphics[width=50mm]{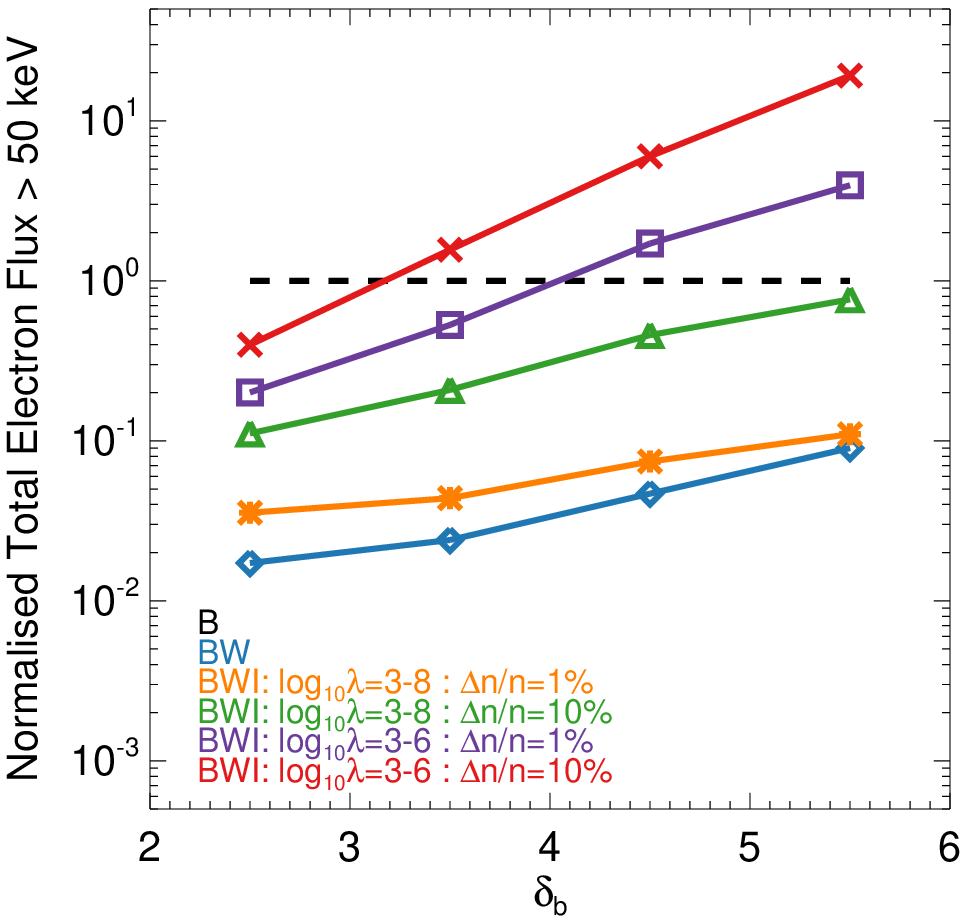}
\caption{\label{fig:nfits}Total number of electrons above $E_\mathrm{C}$,
20 keV and 50 keV (left to right) as a function of the power-law index of the
initial electron distribution $\delta_\mathrm{b}$, obtained from the fitting of
\texttt{f\_thick2.pro} to the simulated footpoint X-ray spectra. The differently
coloured lines indicate the different simulation setups used. The number of
electrons are normalised by those from the fit to the beam-only (B, black line)
X-ray spectrum.}
\end{figure*}

We normalised all fitted results by those found for the B case. They are shown
for the spectral indices $\delta_\mathrm{TT}$ in Figure \ref{fig:delfits} and
total number of electrons $N(>E_\mathrm{C})$ in Figure \ref{fig:nfits}. For the
steepest initial distribution ($\delta_\mathrm{b}=5.5$) all fitted spectral
indices are considerably lower, over 50\% lower for the case with the
strongest density fluctuations. For this level of turbulence in the background
plasma the fitted spectral index is always at least a half of the source index,
indicating the consistent flattening and hardening of the spectrum. The only
exception to this is for the BW simulation with the
hardest source spectrum ($\delta_\mathrm{b}=2.5$) where the fitted index is 20\%
higher. This steeper spectrum is  because the loss of higher energy electrons
from the initially harder distribution (through the generation of Langmuir
waves) dominates over the re-acceleration through the plasma inhomogeneities.

The total number of electrons in the source distribution inferred from the
\texttt{f\_thick2.pro} fits is substantially smaller than the true values (first
panel in Figure \ref{fig:nfits}). The CTTM interpretation of these spectra is a
considerable underestimate (10 to 1\,000 times) of the number of electrons in
the source distribution. It is clear in the mean electron spectrum (Figure
\ref{fig:fspec}) that including the plasma inhomogeneity accelerates electrons
to higher energies, which increases the population at energies well above 20
keV. Therefore we also calculated the number of electrons that the
\texttt{f\_thick2.pro} fit suggests is above 20 and 50 keV in the source
distribution, as shown in the middle and last panels of Figure \ref{fig:nfits}.
At best, with steep spectra and the strongest fluctuations considered here, the
wave scattering can produce a similar number of electrons to those found in the
CTTM case for the number of electrons above 20 keV. For the highest energy
electrons the situation is considerably better with several simulation setups
producing number of electrons $N(>50$keV$)$ similar to or larger than the
B case. For the steepest initial spectrum and strongest fluctuations the
\texttt{f\_thick2.pro} fit overestimates the number of electrons in the source
by over an order-of-magnitude, about a factor of 20. However, it is only
a limited set of conditions ($L^{-1}\geq 10^{-7}$cm$^{-1}$ and
$\delta_\mathrm{b}\ge 3.5$) that produces more high-energy electrons than
\texttt{f\_thick2.pro}.

\section{Discussion and conclusions}
Including Langmuir waves driven by the propagating electron beam causes
major changes in the energy of the electrons and produces substantially
different X-ray spectra. With no, or low levels $L^{-1}<10^{-8}$cm$^{-1}$ of the
density fluctuations in the background plasma, the dominant effect is
wave-particle interactions that decelerate the electrons, which produces a
flatter spectrum and weaker X-ray emission. If these simulated spectra were
assumed to be caused by the CTTM the number of electrons in the source
distribution would be substantially underestimated . With strong
inhomogeneities ($L^{-1}>10^{-7}$cm$^{-1}$) in the background plasma there is
more re-acceleration of the electrons to higher energies, resulting in harder
(flatter, smaller $\delta$) spectra. Interpreting these simulations with the
CTTM produces either a similar amount or an overestimate (we found up to
$\times20$) to the number of electrons in the source distribution. Langmuir waves, if
generated in solar flares, can produce substantial changes in the
flare-accelerated electron distribution. These effects need to be included
for a more reliable interpretation of flare HXR spectra.

We found for $L^{-1}\geq 10^{-7}$cm$^{-1}$ and
$\delta_\mathrm{b}\ge 3.5$, the electron re-acceleration becomes sharply
pronounced, while the low density gradients are insufficient to shift the
energy quickly enough. Indeed, \citet{2012arXiv1211.2587R} showed that the
strongest acceleration is achieved when the relaxation time is close to the time
scale due to the density inhomogeneity. Because we have a fixed upper limit to
the simulation grid $v=115v_\mathrm{T}$ the re-acceleration of electrons in the
hardest source distributions ($\delta < 3.5$) might be lost. We are developing a
full relativistic treatment of Eqs. (\ref{eq:f}) and (\ref{eq:W}) to study
whether the plasma inhomogeneities can have a greater effect for this flatter
spectral domain. 

Our simulations lack wave-wave interactions. The interaction of Langmuir waves
with ion-sound waves has been shown to produce additional electron
re-acceleration \citep{2012A&A...539A..43K}. These simulations had no spatial
dependence and work is under way to investigate their role in the 1D simulations
presented here. The interaction of Langmuir waves with whistler or kinetic
Alf{\'e}n waves \citep[e.g.][]{2010A&A...519A.114B} might also produce
considerable changes to the electron distribution in flares. The density
fluctuations can effectively change the direction of Langmuir waves, and hence
depart from the 1D model, which could limit the application of our simulations.
Recently \citet{2012A&A...544A.148K} have performed a number of 3D
particle-in-cell PIC simulations with initially mono-energetic beams and have
shown that during 3D relaxation a population of electrons appear that has
velocities exceeding those of the injected electrons. While PIC simulations
cannot predict the long-term evolution of these processes as considered here,
the number of accelerated electrons at the stage of plateau formation closely
matches the numbers estimated using 1D quasilinear equations.

\begin{acknowledgements}

This work is supported by a STFC grant ST/I001808/1 (IGH,EPK). Financial support
by the European Commission through the FP7 HESPE network (FP7-2010-SPACE-263086)
is gratefully acknowledged (HASR, EPK).

\end{acknowledgements}


\end{document}